\documentclass[conference]{IEEEtran}

\usepackage{graphicx}
\usepackage{subfigure}
\usepackage{longtable} 
\usepackage{threeparttable}
\usepackage{booktabs}
\usepackage{float}
\usepackage{multirow} 
\usepackage{amsmath}
\usepackage{array}
\usepackage{textcomp}
\usepackage{footnote}
\usepackage{parnotes}
\usepackage{marvosym}
\usepackage{hyperref}
\usepackage{color}
\usepackage[sort]{cite}
\usepackage{float}

\usepackage[textsize=tiny, disable,textwidth=1.35cm]{todonotes}
\newcommand{\Aron}[1]{\todo[color=blue!10, linecolor=black!50]{\textbf{Aron}: #1}}

\newcommand{\se}[1]{\todo[color=orange!10, linecolor=black!50]{\textbf{Scott}: #1}}

\usepackage{pgfplots}
\usepgfplotslibrary{colorbrewer}
\pgfplotsset{compat = 1.14, cycle list/Set1-8} 
\usetikzlibrary{pgfplots.statistics, pgfplots.colorbrewer,pgfplots.dateplot,
	shadows,
	matrix} 
\usepackage{pgfplotstable}
\usepackage{filecontents}

\usepackage{graphicx}
\pgfplotsset{compat=1.14}

\usetikzlibrary{shadows}
\tikzset{
	*|/.style={
		to path={
			(perpendicular cs: horizontal line through={(\tikztostart)},
			vertical line through={(\tikztotarget)})
			-- (\tikztotarget) \tikztonodes
		}
	}
}

\definecolor{blueLine}{RGB}{57,106,177}
\definecolor{blueFill}{RGB}{114,147,203}
\definecolor{redLine}{RGB}{204,37,41}
\definecolor{greenline}{RGB}{0,250,0}
\definecolor{blackLine}{RGB}{0,0,0}
\definecolor{goldLine}{RGB}{160,82,45}
\hypersetup{breaklinks=true}

\begin{document}
	\setlength{\marginparwidth}{1.35cm}
	
	\title{Cyber-Physical Simulation Platform for Security Assessment of Transactive Energy Systems}

\author{\IEEEauthorblockN{Yue~Zhang\IEEEauthorrefmark{1}, Scott Eisele\IEEEauthorrefmark{2}, Abhishek Dubey\IEEEauthorrefmark{2},
Aron Laszka\IEEEauthorrefmark{3},
Anurag~K.~Srivastava\IEEEauthorrefmark{1}}
\IEEEauthorblockA{\IEEEauthorrefmark{1}
\textit{Washington State University}
}
\IEEEauthorblockA{\IEEEauthorrefmark{2}
\textit{Vanderbilt University}
}
\IEEEauthorblockA{\IEEEauthorrefmark{3}
\textit{University of Houston}
}
}
\maketitle
	
	
	\begin{abstract}
	    Transactive energy systems (TES) are emerging as a transformative solution for the problems that distribution system operators face due to an increase in the use of distributed energy resources and rapid growth in scalability of managing active distribution system (ADS). On the one hand, these changes pose a decentralized power system control problem, requiring strategic control to maintain reliability and resiliency for the community and for the utility. On the other hand, they require robust financial markets while allowing participation from diverse prosumers. To support the computing and flexibility requirements of TES while preserving privacy and security, distributed software platforms are required. In this paper, we enable the study and analysis of security concerns by developing Transactive Energy Security Simulation Testbed (TESST), a TES testbed for simulating various cyber attacks. In this work, the testbed is used for TES simulation with centralized clearing market, highlighting weaknesses in a centralized system. Additionally, we present a blockchain enabled decentralized market solution supported by distributed computing for TES, which on one hand can alleviate some of the problems that we identify, but on the other hand, may introduce newer issues. Future study of these differing paradigms is necessary and will continue as we develop our security simulation testbed.

	\end{abstract}

	\begin{IEEEkeywords}
		Cyber-attacks, Transactive Energy Systems, Cyber-Physical Security, Simulation Platform, Testbed.
	\end{IEEEkeywords}

	\section{Introduction}
	
	\begin{figure*}[htp]
		\centering
		\includegraphics[width=\textwidth]{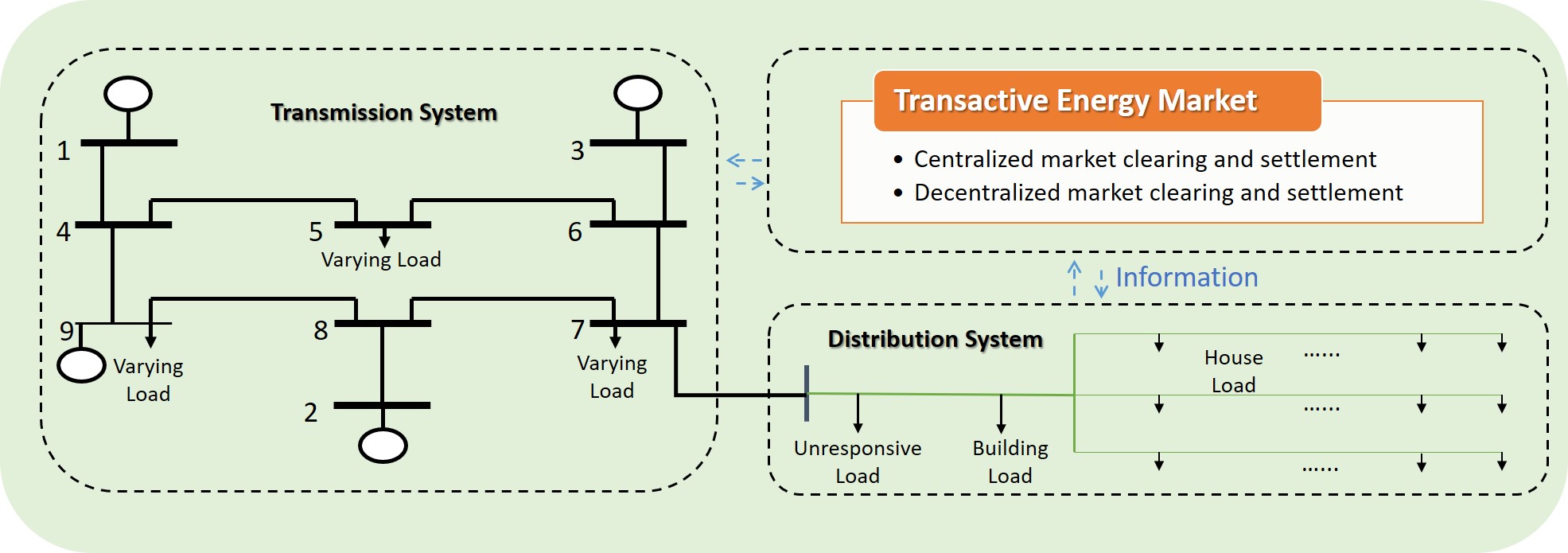}
		\caption{Architecture of TESST.}
		\label{fig:TESSTarch}
	\end{figure*}
	
Transactive energy systems (TES) have emerged as an anticipated outcome of the shift in the electricity industry, moving from centralized, monolithic business models characterized by bulk generation and one-way delivery, toward a decentralized hierarchical model in which end-users can play a more active role in both energy production and consumption \cite{NIST_TE,Gridwise}. In the U.S., 36\% of electricity demand is from single-family houses, which can contribute an even larger share during summer peak due to the usage of air-conditioning~\cite{r1}. The development of smart home devices enables the deployment of TES to provide a more efficient and secure solution. There are a number of well-documented factors contributing to this shift, including the increasing penetration of distributed energy resources, growing number of control variables in the active distribution system, increasing deterioration and fragility of the existing grid, the regulatory and public mandate for environmental awareness, and general social trends toward the democratization of services as exemplified by the ``sharing economy''~\cite{TheGrid}.

TES involving responsive load and distributed generators have received significant attention in the literature. In~\cite{r1}, a transactive control approach is proposed to coordinate heating, ventilation, and air-conditioning systems to reduce load consumption. A distribution locational marginal pricing (DLMP) algorithm is developed to provide a price signal to relax congestion issues in systems with electric vehicles that can act as prosumers in \cite{r6}. A coordination method that can manage energy imbalance problems considering thermostatically controllable load is presented in \cite{r7}. An automated decentralized control scheme is introduced to provide ancillary and demand response services in \cite{r9}. A TES that could maximize resource utilization and balance demand and supply is proposed in \cite{r10,r11}. A TES that allows direct control of unit consumption through an aggregator is introduced in \cite{r12}. A double-auction market scheme that utilizes transactive controllers to operate the distribution system is designed in~\cite{r13}.

The transactive market can be implemented with multiple possible architectures. Most of the architecture will be hierarchical and can have alternate architecture at a different voltage level. Prosumers can coordinate with aggregators or campus grid, and aggregators/campus grid can coordinate with distribution system operators. DSO can coordinate with transmission system operators to optimize resources in the best possible way. We consider two different architectures: 1) hierarchical with centralized market clearing and 2) hierarchical with a mix of centralized and distributed market clearing enabled by enhanced communication. First architecture is well explored in literature but not much for security analysis. Information exchange between prosumers and a system operator or aggregators happens through a large number of distributed edge-computing and Internet of Things (IoT) devices.
TES communication is conducted with digital infrastructure and requires interfacing with edge-devices, which have possible vulnerabilities and attacks especially with financial interest motives. The main actors are the consumers, which comprise primarily residential loads and prosumers who also have distributed energy resources (DERs), such as rooftop solar batteries or flexible loads capable of demand/response. Additionally, a distribution system operator (DSO) manages the network with possible additional interface with microgrid operator or campus grid operators and with prosumers directly or through aggregators. 

For second architecture, such installations are equipped with an advanced metering infrastructure consisting of TE-enabled smart meters. In addition to the standard functionalities of smart meters (i.e., the ability to measure line voltages, power consumption and net metering, and to communicate these to the distribution system operator), TE-enabled smart meters are capable of communicating with other smart meters, have substantial onboard computational resources, and are able to access the Internet and cloud computing services as needed. Examples of such installations include the well-known Brooklyn Microgrid Project, \cite{BrooklynMicrogrid} and the Sterling Ranch learning community (currently under development) \cite{SterlingRanch}.
	
	The research community is increasingly advocating the use of distributed ledgers in TES, including our earlier work \cite{laszka2018transax}. Blockchain technology enables the digital representation of energy and financial assets and their secure transfer from one set of parties to another. By design, the security of this value transfer is guaranteed by the interaction protocol itself and obviates the need for trusted transaction intermediaries. The execution of smart contracts (i.e., code that captures the market logic and participants' roles) is automated and guaranteed~\cite{underwood2016blockchain,mavridou2019verisolid}. Additionally, the blockchain constitutes an immutable, complete, and fully auditable record of all transactions that have occurred within the system. These properties ensure market transparency, as well as the availability of a detailed market load profile and grid utilization data. 
	
    {\bf Problem:} In this paper, we specifically consider the problem of security in a transactive energy system. Unlike traditional power grid operation, the participation of the prosumers at the edge raises several concerns. The first concern is privacy: if private data is stored in a way that is easily accessible to unauthorized entities,  it can leak private information. Consider that the transaction level data can provide much greater insights into a prosumer’s behavior compared to smart meter
data \cite{Laszka:2017:PPS:3131542.3131562}. Similarly, the market is an integral part of a TES. Hence, it is important that the market remains fair and cannot be manipulated. Consider the problem of a set of prosumers promising to supply energy at a lower bid and then choosing not to supply the power at the scheduled time.

    {\bf Contributions:} 
To systematically study these adverse scenarios, we must have access not only to a power system simulator but also to a system that can simulate market mechanisms, including distributed ledgers if they are integrated. In this paper, we build on the transactive energy simulation platform (TESP) developed by the Pacific Northwest National Lab (PNNL). We extended the TESP platform to study security scenarios by incorporating different attacks. We call this testbed the \emph{Transactive Energy Security Simulation Testbed} (TESST). We highlight weaknesses in TES with a hierarchical and centralized market clearing system. Then, we discuss how some of these problems can be alleviated by the use of a decentralized market solution based on our earlier work \cite{laszka2018transax}. Future work includes detailed security analysis for the ledger enabled TESP and integrating a suite of attack scenarios that can be used by the research community.


    


	\section{TESST: TES Testbed for Cyber Attack Simulation}

	Transactive Energy Security Simulation Testbed (TESST) is built upon the Transactive Energy Simulation Platform (TESP) by PNNL and TRANSAX designed by Vanderbilt University\cite{laszka2018transax,laszka2017providing}. The TESP platform is interfaced with Network Simulator 3 (NS3) to simulate cyber attacks on the TES\cite{Krishnan_2018,Arman_2018} (please see Figure~\ref{fig:TESSTarch}).

	\subsection{Physical System}

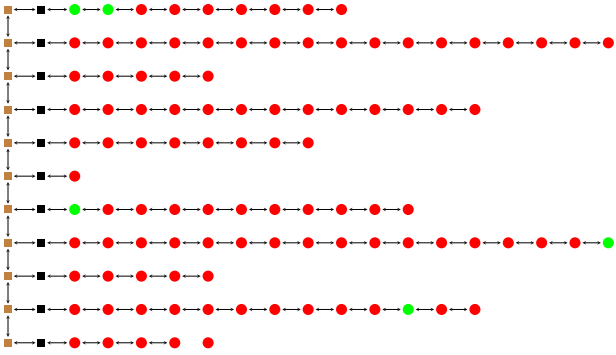
\begin{figure}[t]
 \centering
\resizebox{0.45\textwidth}{!}{%
 \begin{tikzpicture}[rotate=270,font=\tiny,
   oc/.style={fill=black,rectangle,minimum size=0.01cm,font=\tiny},
     feeder/.style={fill=brown,rectangle,minimum size=0.005cm,font=\tiny},
   Producer/.style={fill=green,circle,minimum size=0.01cm},
     Consumer/.style={fill=red,circle,minimum size=0.01cm},
   Connection/.style={<->, >=stealth, shorten <=0.05cm, shorten >=0.05cm}]
 \draw node[oc] (oc1) at (-5,0){};
 \draw node[oc] (oc2) at (-4,0){};
 \draw node[oc] (oc3) at (-3,0){};
 \draw node[oc] (oc4)  at (-2,0){};
 \draw node[oc](oc5)  at (-1,0){};
 \draw node[oc] (oc6)  at (0,0){};
 \draw node[oc] (oc7) at (1,0){};
 \draw node[oc] (oc8)  at (2,0){};
 \draw node[oc] (oc9) at (3,0){};
 \draw node[oc] (oc10) at (4,0){};
 \draw node[oc] (oc11) at (5,0){};

 \draw node[feeder] (feeder1) at (-5,-1){};
 \draw node[feeder] (feeder2) at (-4,-1){};
 \draw node[feeder] (feeder3) at (-3,-1){};
 \draw node[feeder] (feeder4)  at (-2,-1){};
 \draw node[feeder](feeder5)  at (-1,-1){};
 \draw node[feeder] (feeder6)  at (0,-1){};
 \draw node[feeder] (feeder7) at (1,-1){};
 \draw node[feeder] (feeder8)  at (2,-1){};
 \draw node[feeder] (feeder9) at (3,-1){};
 \draw node[feeder] (feeder10) at (4,-1){};
 \draw node[feeder] (feeder11) at (5,-1){};

 \draw [Connection] (feeder1) to (feeder2);
 \draw [Connection] (feeder2) to (feeder3);
 \draw [Connection] (feeder3) to (feeder4);
 \draw [Connection] (feeder4) to (feeder5);
 \draw [Connection] (feeder5) to (feeder6);
 \draw [Connection] (feeder6) to (feeder7);
 \draw [Connection] (feeder7) to (feeder8);
 \draw [Connection] (feeder8) to (feeder9);
 \draw [Connection] (feeder9) to (feeder10);
 \draw [Connection] (feeder10) to (feeder11);

\draw [Connection] (feeder1) to (oc1);
\draw [Connection] (feeder2) to (oc2);
\draw [Connection] (feeder3) to (oc3);
\draw [Connection] (feeder4) to (oc4);
\draw [Connection] (feeder5) to (oc5);
\draw [Connection] (feeder6) to (oc6);
\draw [Connection] (feeder7) to (oc7);
\draw [Connection] (feeder8) to (oc8);
\draw [Connection] (feeder9) to (oc9);
\draw [Connection] (feeder10) to (oc10);
\draw [Connection] (feeder11) to (oc11);

 \foreach \pos in {1,2} {
   \node [Producer] (p10\pos)at (-5,\pos) {};
 }

 \foreach \pos in {3,4,5,6,7,8,9} {
   \node [Consumer] (c10\pos)at (-5,\pos) {};
 }

 \foreach \pos in {1,2,3,4,5,6,7,8,9,10,11,12,13,14,15,16,17} {
   \node [Consumer] (c20\pos)at (-4,\pos) {};
 }

 \foreach \pos in {1,2,3,4,5} {
   \node [Consumer] (c30\pos)at (-3,\pos) {};
 }

 \foreach \pos in {1,2,3,4,5,6,7,8,9,10,11,12,13} {
   \node [Consumer] (c40\pos)at (-2,\pos) {};
 }

 \foreach \pos in {1,2,3,4,5,6,7,8} {
   \node [Consumer] (c50\pos)at (-1,\pos) {};
 }

 \foreach \pos in {1} {
   \node [Consumer] (c60\pos)at (0,\pos) {};
 }

 \foreach \pos in {1} {
   \node [Producer] (p70\pos)at (1,\pos) {};
 }

 \foreach \pos in {2,3,4,5,6,7,8,9,10,11} {
   \node [Consumer] (c70\pos)at (1,\pos) {};
 }

 \foreach \pos in {17} {
   \node [Producer] (p80\pos)at (2,\pos) {};
 }

 \foreach \pos in {1,2,3,4,5,6,7,8,9,10,11,12,13,14,15,16} {
   \node [Consumer] (c80\pos)at (2,\pos) {};
 }

 \foreach \pos in {1,2,3,4,5} {
   \node [Consumer] (c90\pos)at (3,\pos) {};
 }

 \foreach \pos in {11} {
   \node [Producer] (p100\pos)at (4,\pos) {};
 }

 \foreach \pos in {1,2,3,4,5,6,7,8,9,10,12,13} {
   \node [Consumer] (c100\pos)at (4,\pos) {};
 }

 \foreach \pos in {1,2,3,4,5} {
   \node [Consumer] (c110\pos)at (5,\pos) {};
 }

 \draw [Connection] (oc1) to (p101);
 \draw [Connection] (p101) to (p102);
 \draw [Connection] (p102) to (c103);
 \draw [Connection] (c103) to (c104);
 \draw [Connection] (c104) to (c105);
 \draw [Connection] (c105) to (c106);
 \draw [Connection] (c106) to (c107);
 \draw [Connection] (c107) to (c108);
 \draw [Connection] (c108) to (c109);
 
\draw [Connection] (oc2) to (c201);
\draw [Connection] (c201) to (c202);
\draw [Connection] (c202) to (c203);
\draw [Connection] (c203) to (c204);
\draw [Connection] (c204) to (c205);
\draw [Connection] (c205) to (c206);
\draw [Connection] (c206) to (c207);
\draw [Connection] (c207) to (c208);
\draw [Connection] (c208) to (c209);
\draw [Connection] (c209) to (c2010);
\draw [Connection] (c2010) to (c2011);
\draw [Connection] (c2011) to (c2012);
\draw [Connection] (c2012) to (c2013);
\draw [Connection] (c2013) to (c2014);
\draw [Connection] (c2014) to (c2015);
\draw [Connection] (c2015) to (c2016);
\draw [Connection] (c2016) to (c2017);

\draw [Connection] (oc3) to (c301);
\draw [Connection] (c301) to (c302);
\draw [Connection] (c302) to (c303);
\draw [Connection] (c303) to (c304);
\draw [Connection] (c304) to (c305);

\draw [Connection] (oc4) to (c401);
\draw [Connection] (c401) to (c402);
\draw [Connection] (c402) to (c403);
\draw [Connection] (c403) to (c404);
\draw [Connection] (c404) to (c405);
\draw [Connection] (c405) to (c406);
\draw [Connection] (c406) to (c407);
\draw [Connection] (c407) to (c408);
\draw [Connection] (c408) to (c409);
\draw [Connection] (c409) to (c4010);
\draw [Connection] (c4010) to (c4011);
\draw [Connection] (c4011) to (c4012);
\draw [Connection] (c4012) to (c4013);

\draw [Connection] (oc5) to (c501);
\draw [Connection] (c501) to (c502);
\draw [Connection] (c502) to (c503);
\draw [Connection] (c503) to (c504);
\draw [Connection] (c504) to (c505);
\draw [Connection] (c505) to (c506);
\draw [Connection] (c506) to (c507);
\draw [Connection] (c507) to (c508);

\draw [Connection] (oc6) to (c601);

\draw [Connection] (oc7) to (p701);
\draw [Connection] (p701) to (c702);
\draw [Connection] (c702) to (c703);
\draw [Connection] (c703) to (c704);
\draw [Connection] (c704) to (c705);
\draw [Connection] (c705) to (c706);
\draw [Connection] (c706) to (c707);
\draw [Connection] (c707) to (c708);
\draw [Connection] (c708) to (c709);
\draw [Connection] (c709) to (c7010);
\draw [Connection] (c7010) to (c7011);

\draw [Connection] (oc8) to (c801);
\draw [Connection] (c801) to (c802);
\draw [Connection] (c802) to (c803);
\draw [Connection] (c803) to (c804);
\draw [Connection] (c804) to (c805);
\draw [Connection] (c805) to (c806);
\draw [Connection] (c806) to (c807);
\draw [Connection] (c807) to (c808);
\draw [Connection] (c808) to (c809);
\draw [Connection] (c809) to (c8010);
\draw [Connection] (c8010) to (c8011);
\draw [Connection] (c8011) to (c8012);
\draw [Connection] (c8012) to (c8013);
\draw [Connection] (c8013) to (c8014);
\draw [Connection] (c8014) to (c8015);
\draw [Connection] (c8015) to (c8016);
\draw [Connection] (c8016) to (p8017);

\draw [Connection] (oc9) to (c901);
\draw [Connection] (c901) to (c902);
\draw [Connection] (c902) to (c903);
\draw [Connection] (c903) to (c904);
\draw [Connection] (c904) to (c905);

\draw [Connection] (oc10) to (c1001);
\draw [Connection] (c1001) to (c1002);
\draw [Connection] (c1002) to (c1003);
\draw [Connection] (c1003) to (c1004);
\draw [Connection] (c1004) to (c1005);
\draw [Connection] (c1005) to (c1006);
\draw [Connection] (c1006) to (c1007);
\draw [Connection] (c1007) to (c1008);
\draw [Connection] (c1008) to (c1009);
\draw [Connection] (c1009) to (c10010);
\draw [Connection] (c10010) to (p10011);
\draw [Connection] (p10011) to (c10012);
\draw [Connection] (c10012) to (c10013);

\draw [Connection] (oc11) to (c1101);
\draw [Connection] (c1101) to (c1102);
\draw [Connection] (c1102) to (c1103);
\draw [Connection] (c1103) to (c1104);

 \end{tikzpicture}
 }
 \caption{Feeder diagram. Brown nodes are feeder junctions, numbered 1 to 11 from top to bottom.  Black nodes are the overcurrent relays, which ensure that the total power flowing in and out of the feeder is below 20 kW. The green nodes are the junction points for the producers ($5$), and the red nodes are junction points for the consumers ($97$). There are $102$ prosumers in total.}
 \label{fig:feeder}
 \end{figure}

	The physical system includes a modified IEEE 9 bus system and a distribution feeder connected at bus 7. The transmission system is modeled in PyPower and includes four fossil based generating units. The cost of the unit connected at bus 9 is higher than the other units. There are also three varying loads connected at buses 5, 7 and 9. The distribution system is modeled with Gridlab-D and EnergyPlus. In this distribution feeder, the 1.3MW unresponsive load is connected at 12.47KV voltage level. The building load is also connected to that voltage level through a 12.47kV/480V transformer. There are also 30 houses that are equipped with PV panels and Heating, Ventilation, and Air Conditioning (HVAC) systems, which are connected to the feeder through a 7200V/120V transformer. Moreover, a microgrid that containing $102$ homes across $11$ feeders ($5$ producers and $97$ consumers) is also built. The feeder structure and its safety limits are plotted in Fig. ~\ref{fig:feeder}.
	
   
    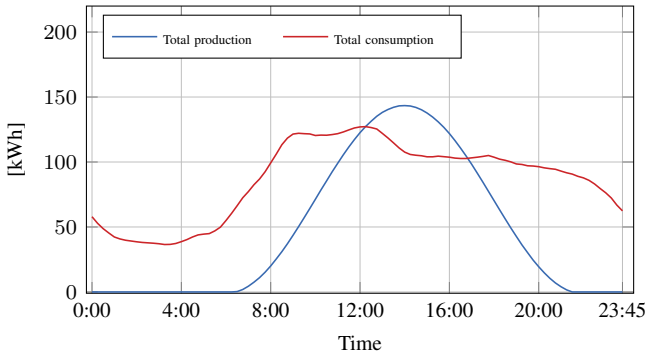
\begin{figure}[t]
\centering
\begin{tikzpicture}
\begin{axis}[
  font=\footnotesize,
  width=\columnwidth,
  height=0.61\columnwidth,
  ymin=-1,
 legend columns=3, 
  legend style={font=\fontsize{5}{6}\selectfont,cells={align=left},text width=4.3em,text height=1.5ex,text depth=.5ex,row sep=0.1em},
  ymax=220,
grid=both,
    grid style={line width=.1pt, draw=gray!10},
    major grid style={line width=.2pt,draw=gray!50},
  xmin=-1,
  xmax=97,
  legend pos=north west,
  xlabel=Time,
  ylabel={[kWh]},
  ytick={0, 50, 100, 150, 200},
  xtick={0, 16, 32, 48, 64, 80, 95},
  xticklabels={0:00, 4:00, 8:00, 12:00, 16:00,  20:00, 23:45},
]
\addplot[no markers, solid, blueLine, semithick] table[x expr=\coordindex, y=TotalProduction, comment chars={\%}] {diagrams/TotalProductionConsumption.csv};
\addlegendentry{Total production};
\addplot[no markers, solid, redLine, semithick] table[x expr=\coordindex, y=TotalConsumption, comment chars={\%}] {diagrams/TotalProductionConsumption.csv};
\addlegendentry{Total consumption};
\end{axis}
\end{tikzpicture}
\caption{Load profile (i.e., total consumption) and generation profile (i.e., total production) in kWh per 15 minute interval aggregated across the microgrid.}
\label{fig:profile}
\end{figure}

	\subsection{Transactive Energy Market}
	
	Two trading options are designed for the TESST. The transactive energy market has one option with a centralized market clearing and settlement and the second has decentralized market clearing and settlements. 
	
	\subsubsection{Centralized Market Option}
	The centralized option utilizes a double-auction market mechanism and aims to provide a trading platform for prosumers in both transmission and distribution systems. Both suppliers and consumers submit their bids to the TES management platform, and the market uses the auction to determine the clearing price which is published through the network to all participants. 
	
	\begin{figure}[htp]
		\centering
		\includegraphics[width=3.3in]{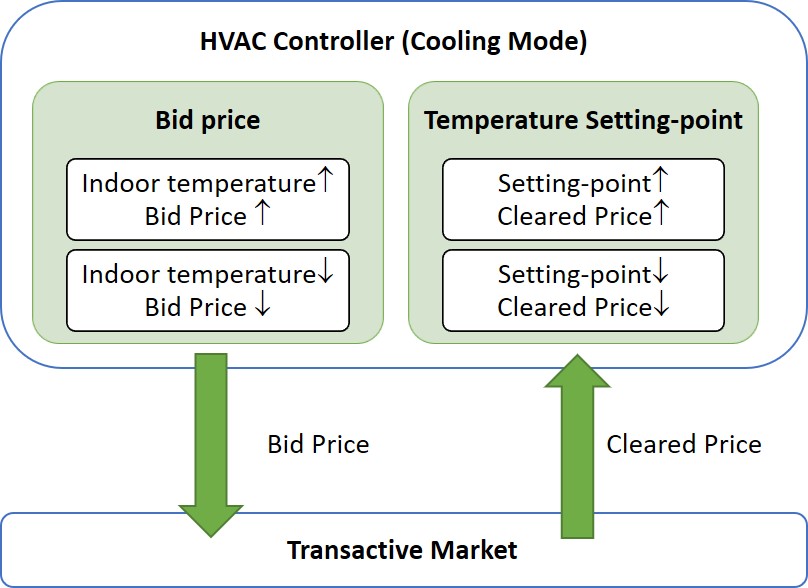}
		\caption{Smart HVAC controller mechanism in cooling mode.}
		\label{HVAC Controller in cooling mode}
	\end{figure}

	The consumer bids are from smart HVAC controllers, which will adjust bid price and quantity based on the most recent cleared price and the average and standard deviation of the cleared price over the preceding 24 hours. They can also adjust the temperature settings to earn more savings. For example, the adjustment process under cooling mode is described in Figure~\ref{HVAC Controller in cooling mode}. The smart HVAC controller will adjust bid based on current temperature and adjust temperature setting based on current cleared price using the following two equations:
	\begin{align}
	T_{\textnormal{Set}}=T_{\textnormal{Target}}+\dfrac{(P_{\textnormal{Clear}}-P_{\textnormal{Mean}})\cdot |T_{max/min}|}{\sigma_T \cdot \sigma_P}\\
	P_{\textnormal{Bid}}=P_{\textnormal{Mean}}+\dfrac{(T_{\textnormal{Current}}-T_{\textnormal{Target}})\cdot \sigma_T \cdot \sigma_P}{|T_{max/min}|}
	\end{align}
	where $T_{\textnormal{Set}}$ is the new adjusted temperature setpoint; $T_{\textnormal{Target}}$ is the target temperature setpoint; $P_{\textnormal{Clear}}$ is the received cleared price; $P_{\textnormal{Mean}}$ is the average price over the last 24 hours; $T_{max}$ and $T_{min}$ are the maximum and minimum acceptable temperature; $\sigma_T$ and $\sigma_P$ are the standard deviation of temperature and price, respectively; $P_{\textnormal{Bid}}$ is the bid price from HVAC controller; $T_{\textnormal{Current}}$ is the current air temperature.

	If the received cleared price is higher than the threshold, the temperature setpoint is moved to a higher value to decrease electricity consumption. Otherwise, the setpoint is moved to a lower value to gain a higher comfort level. The bid price is determined by the current air temperature. For example, on a hot day, if the temperature is very high, the HVAC controller will select a high bid price to increase the chance of acceptance for its bid. The smart HVAC controllers also utilize the average price information to adjust consumption patterns more efficiently to avoid frequently changing the setting for short-term variation.

	\subsubsection{Decentralized Market Option}
	\label{decentralized-market}
	
	The decentralized option utilizes a decentralized middleware called Resilient Information Architecture Platform for Smart Grid (RIAPS) \cite{isorc2017} to create a framework for decentralized energy trading. 
	
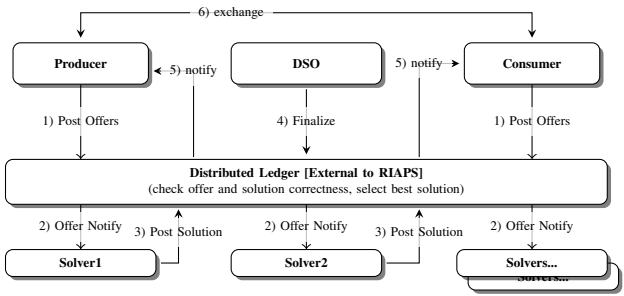
\begin{figure}[t]
\centering
\begin{tikzpicture}[x=1.5cm, y=0.9cm, font=\tiny,
  Component/.style={fill=white, draw, align=center, rounded corners=0.1cm, drop shadow={shadow xshift=0.05cm, shadow yshift=-0.05cm, fill=black}},
  Connection/.style={<->, >=stealth, shorten <=0.06cm, shorten >=0.06cm},
  Label/.style={midway, align=center, fill=white, fill opacity=0.75, text opacity=1}
]
  
\node [Component, minimum width=2cm] (solver1) at (-2, -0.2) {\textbf{Solver1}};
\node [Component, minimum width=2cm] (solver2) at (0, -0.2) {\textbf{Solver2}};
\node [Component, minimum width=2cm] (solvers2) at (2.1, -0.4) {\textbf{Solvers...}};
\node [Component, minimum width=2cm] (solvers) at (2, -0.2) {\textbf{Solvers...}};

\node [Component, align=center, minimum width=8cm] (sc) at (0, 1) {\textbf{Distributed Ledger [External to RIAPS]}\\(check offer and solution correctness, select best solution)};

\node [Component, minimum width=1.8cm, minimum height=0.55cm] (producer) at (-2, 2.75) {\textbf{Producer}};
\node [Component, minimum width=2cm, minimum height=0.55cm] (DSO) at (0, 2.75) {\textbf{DSO}};
\node [Component, minimum width=1.8cm, minimum height=0.55cm] (consumer) at (2, 2.75) {\textbf{Consumer}};

\draw[<-,*|]  (sc.north) to node [Label] {1) Post Offers} (producer.south) ;
\draw[<-,*|]  (sc.north) to node [Label] {1) Post Offers} (consumer.south) ;

\draw[->,*|]  (sc.south) to node [Label] {2) Offer Notify} (solver1.north) ;
\draw[->,*|]  (sc.south) to node [Label] {2) Offer Notify} (solver2.north) ;
\draw[->,*|]  (sc.south) to node [Label] {2) Offer Notify} (solvers.north) ;

\draw[Connection, ->] (solver1.east) -| node [Label, yshift=0.4cm] {3) Post Solution} ([xshift=-1.7cm]sc.south);
\draw[Connection, ->] (solver2.east) -| node [Label, yshift=0.4cm] {3) Post Solution} ([xshift=1.5cm]sc.south);

\draw [Connection, ->] (DSO) -- node [Label] {4) Finalize} (sc.north);

\draw [Connection, <-] (consumer.west) -| node [Label] {5) notify} (1,1.3);
\draw [Connection, <-] ([yshift=-.1cm]producer.east) -| node [Label] {5) notify} (-1,1.3);

\draw [Connection] (producer.north) |- node [Label, xshift=2cm] {6) exchange}(1,3.5) -| (consumer.north);

\end{tikzpicture}
    \caption{Data flow between actors of in Transactive Energy application. 
        }
    \label{fig:transax}
\end{figure}

	The actors and high-level data-flow of this platform can be seen in Figure \ref{fig:transax}. The typical workflow begins with producers and consumers of power (1) posting offers to the distributed ledger, offering to sell or buy energy for a time interval in the future. In \cite{laszka2018transax}, we used Ethereum as the ledger. Solvers monitor the ledger, and when offers are posted, they (2) use an algorithm to match buyers to sellers. This match is (3) posted to the ledger. The solution for an interval may be updated until it is (4) finalized by the DSO. At this point, producers and consumers are notified and will (6) exchange the amount of power for which they were matched. RIAPS was used to provide inter-actor communication, management services, and time-synchronization for the actors to begin the transfer of power at the right time.

	The trading scenarios that we consider involve consumers and prosumers that participate in a local P2P energy trading market by posting offers to sell produced energy or to buy and consume energy in a future time interval. An offer consists of the quantity of energy being bought or sold, the time interval in which the trade is to be delivered, and possibly a reservation price, i.e., the maximum (or respectively, minimum) price at which the buyer (or respectively, seller) is willing to trade.
	
	We assume that each participant predicts their future power production and consumption (e.g., based on historical data) and does so prior to trading on the market. 
	Moreover, each participant is represented by an automated trading agent that strategically posts offers to the TES management platform (TMP) based on these predictions and the participant's personal trading goals.
	
	In the simplest trading scenario, the DSO sets the price $p$ per kWh for the local market; $p$ is the price paid by any buyer and received by any seller, including the DSO. The DSO can then dynamically adjust the price $p$ to affect the market efficiency, which is evaluated as the number of local transactions vs. energy demand met from a bulk supplier. Another scenario includes a fully dynamic market where all sellers, including the DSO, post offers that include a reservation price. Each consumer then picks a selling offer on a first-come, first served basis. An extension of this scenario involves double auctions where both selling and buying offers are posted to the TMP, which executes an automated, regulator-approved market clearing algorithm as an immutable smart contract on the TMP's blockchain system. This algorithm selects the clearing price of $p$ within each time interval.
    
\begin{figure}[t]
\begin{tikzpicture}
\begin{axis}[
  font=\footnotesize,
  width=\columnwidth,
  height=0.61\columnwidth,
  ymin=2865,
  ymax=2975,
  ytick={2875,2900,2925,2950},
  xtick={2,3,5,7,10,13},
  legend pos=north west,
  xlabel={Prosumer prediction window [time intervals]},
  ylabel={Total energy traded [kWh]},
grid=both,
    grid style={line width=.1pt, draw=gray!10},
    major grid style={line width=.2pt,draw=gray!50},
]
\addplot[mark=o, color=blueLine, only marks, thick] table[x=predictionWindow, y=withoutBattery, comment chars={\%}, col sep=comma] {diagrams/total-energy-traded.csv};
\addlegendentry{without battery};
\addplot[mark=+, color=redLine, only marks, thick] table[x=predictionWindow, y=withBattery, comment chars={\%}, col sep=comma] {diagrams/total-energy-traded.csv};
\addlegendentry{with battery};
\end{axis}
\end{tikzpicture}
\caption{Total amount of energy traded in the entire microgrid with and without batteries, for various prediction window lengths.}
\label{fig:multihorizon}
	\centering
        \includegraphics[width=\columnwidth]{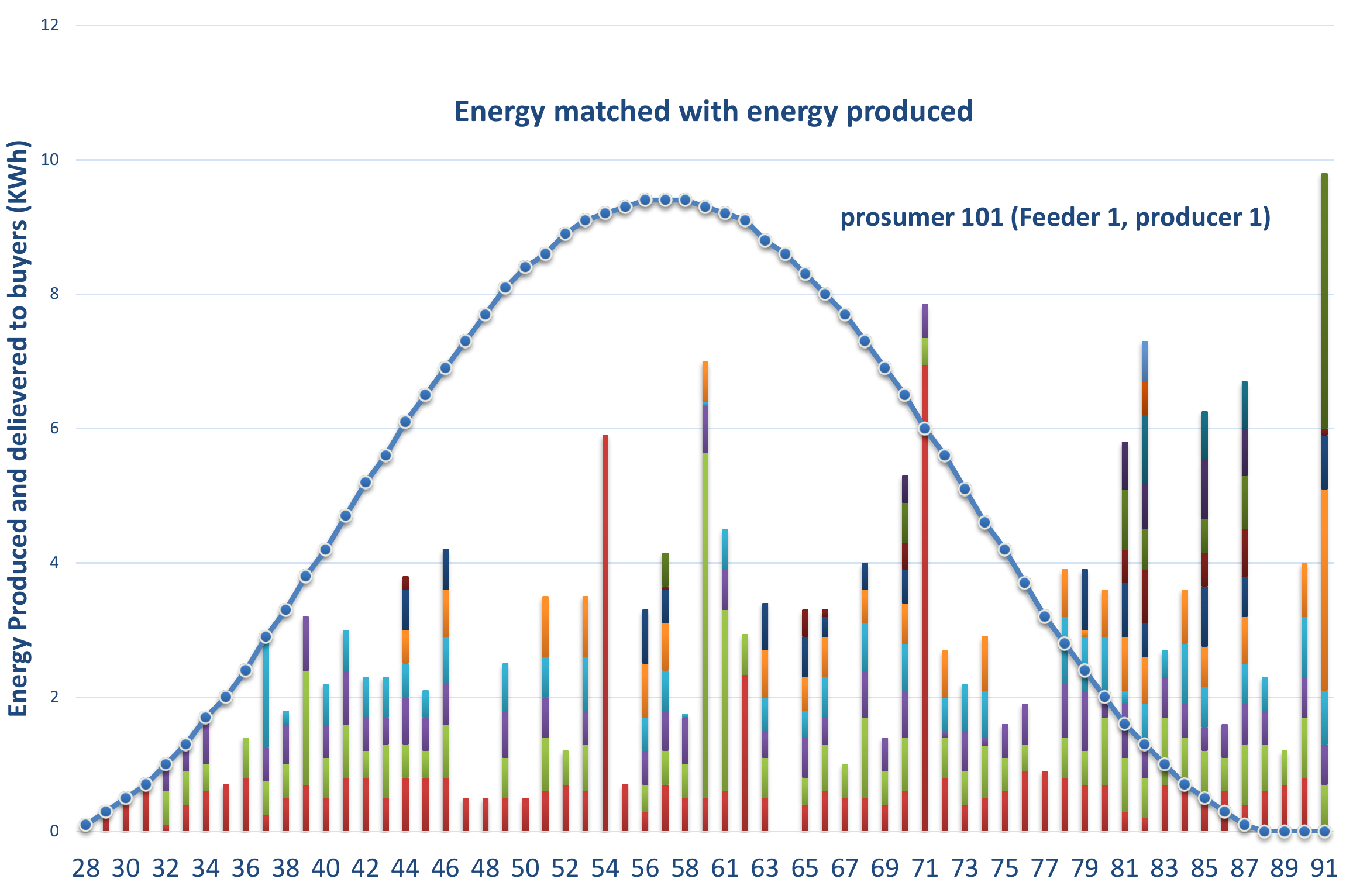}
	    \caption{Energy generated in each interval (blue line) and energy traded to a set of consumers in each interval (vertical bars) for the first prosumer of the first feeder. The stacked colors show the different consumers that were matched with the prosumer in each interval (note that the same color across multiple intervals does not necessarily mean the same consumer). When the energy traded exceeds the generation, the excess is drawn from the battery.}
	\label{fig:allocate1NRG}
    \end{figure}

    One of the innovative capabilities of the decentralized trading platform is the ability to specify multiple time intervals in selling offers, enabling the integration of battery systems for delaying the sale or purchase of energy. Figure \ref{fig:multihorizon} shows the total energy traded for different tests. We varied the prediction window for the participants from $2$ to $13$. That is, in each interval, the participants submitted offers starting from the next $1$ to $12$ intervals (the current interval is always counted in the prediction window). The experiment simulated the whole day from the first interval starting at 0:00 (12:00 AM) to the $95^{th}$ interval ending at 23:59 (11:59 PM). As expected, increasing the prediction window with batteries improves performance, and without batteries has no effect on the total amount of energy traded. This is because any production must be dispatched within one-time interval, so the solver cannot optimize energy usage across multiple intervals even if future offers are available.
	
	An example execution run of the system is shown in 
	Figure~\ref{fig:allocate1NRG}. This figure shows the energy matched per interval for the first prosumer of the first feeder (Figure~\ref{fig:feeder}). 

	\subsection{Network Simulator}
	
\begin{figure}[htp]
		\begin{center}
			\includegraphics[width=\columnwidth]{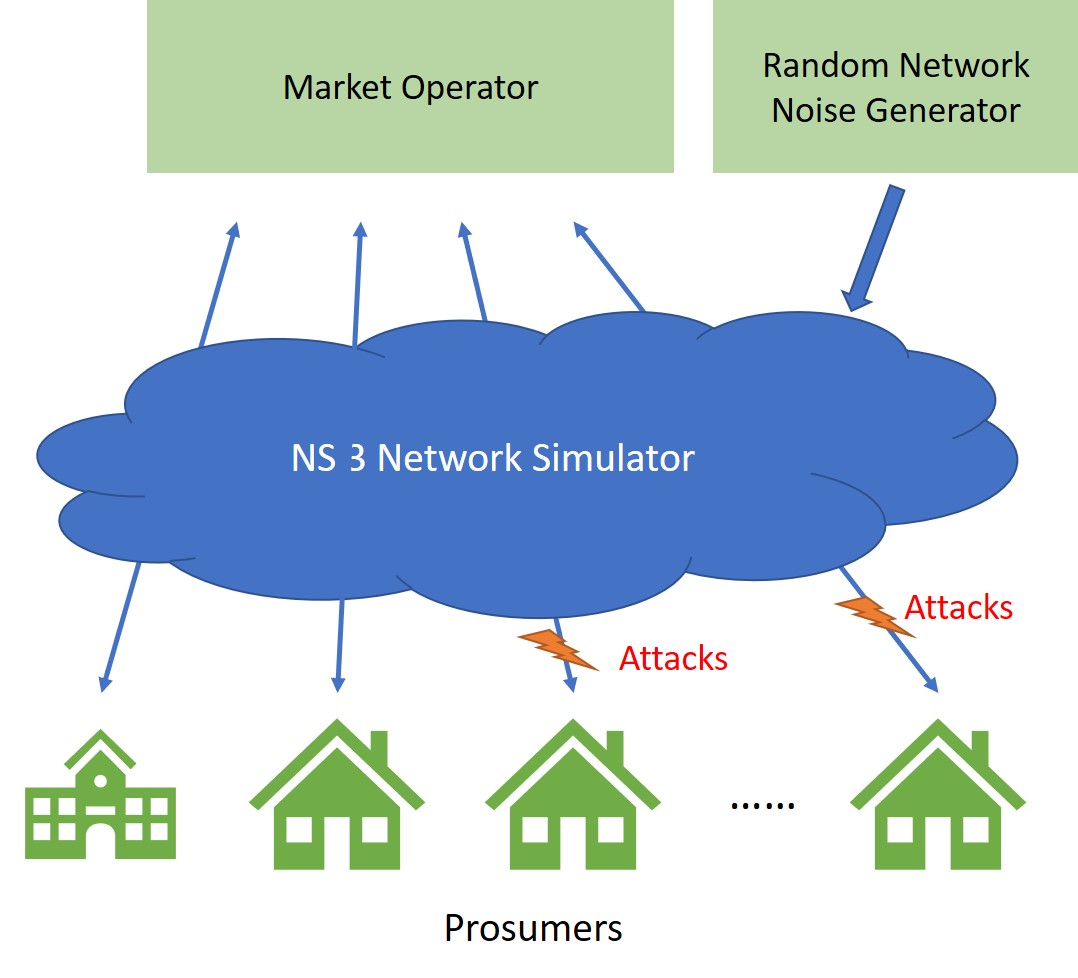}
			\caption{Network simulator for TESST}
			\label{Network simulator for TESST}
		\end{center}
	\end{figure}

	The network simulator, as shown in Figure~\ref{Network simulator for TESST}, creates communication channels among the prosumers and enables the simulation of cyber attacks. The network simulator is built using the tap-bridge module of NS3; here, every container has a bridge and is linked to a tap device in NS3. All tap devices communicate through a virtual wireless network. Besides, random network traffic has been added to the simulation to mimic the real world network situation. The purpose of this random noise is to test if the data analytics implemented in this paper can detect the attacks in the most randomized dataset since the real datasets would also contain such random noise. The randomness in the dataset is equivalent to that of a workstation in a smart grid testbed, and it consists of web traffic and system updates. As the simulation begins, the communication between various components is collected using a packet sniffer tool. The collected information includes IP addresses, port numbers, length of packets, protocol, and number of bytes sent every five minutes.

	\section{Cyber Threat and Security Analysis}
	
    A centralized trading platform may be exposed to a variety of cyber-threats and privacy issues. Some attackers seek financial gain through network-based attacks, and they will manipulate the controllers to profit. Some attackers aim to disturb the operation of the TES. Similar to the notable cyber-attack against Ukrainian power systems in December 2015~\cite{lee2016analysis,zetter2016inside}, attackers can inject malware into the market operation system and manipulate settings, such as DLMP limits or clearing time interval. An attacker could use a malicious channel to eavesdrop on the power system and can also steal critical information. Through stealing critical information from both the market operator and prosumers, attackers could devise a sophisticated targeted attack. They can also conduct Denial-of-Service (DOS) attacks that aim to cause a lack of availability of information, updates, prices, and resources. In contrast to the market operation system, individual smart HVAC controllers do not necessarily employ strong security mechanisms. Therefore, compromise a large number of smart HVAC controllers either for financial gain or to damage the system by sending massively manipulated bids to the market operator.
	
	During the simulation of TESST, the prosumers will submit bids to buy or sell electricity and the market operator will collect those bids and produce a clearing price. This information is crucial for the operation of a TES. If this information is compromised, the operation of the transactive system may be impacted at multiple levels. 

\begin{figure}[htp]
		\centering
		\includegraphics[width=\columnwidth]{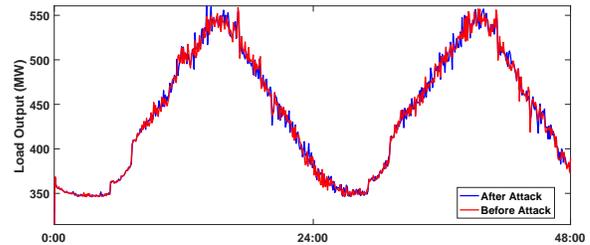}
		\caption{Demand curve changes due to attacks aim for profits}
		\label{Demand curve changes due to attacks aim for profits}
	\end{figure}

Here we explore two possible attacks that assume an attacker is able to manipulate the bid price and quantity\se{I'm not sure how reasonable this assumption is. But I guess its fine for now}\Aron{I think it's fine}. In the first scenario, the attacker seeks personal benefit \se{some implementations of double auction are resistant to this kind of attack. It's not really a centralized market problem. wikipedia double auction truthfulness} by reducing the bid price and quantity by 50\%. \se{how do they benefit from this attack?} If only a limited number of prosumers launch such an attack, it will be difficult to detect because the impact on the total demand curves is small as seen if Figure~\ref{Demand curve changes due to attacks aim for profits}. 
	
\begin{figure}[htp]
		\centering
		\includegraphics[width=\columnwidth]{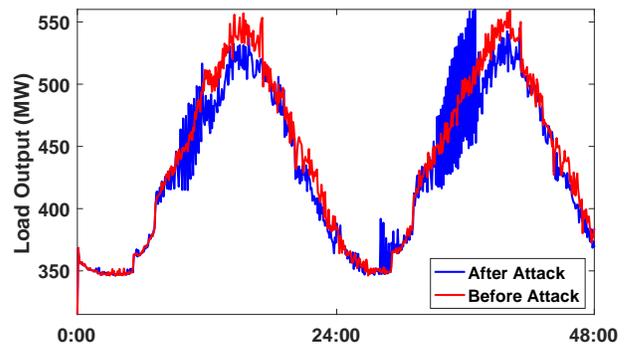}
		\caption{Demand curve changes due to attacks aim for disturbing system operation}
		\label{Demand curve changes due to attacks aim for disturbing system operation}
	\end{figure}

	In the second scenario, the attacker aims to disturb the operation of the transactive market, which can be done by changing the prosumers' bids to arbitrarily high or low values. Such drastic modifications will lead to significant changes in clearing price, the operation of smart HVAC controllers, and the overall demand as shown in Figure~\ref{Demand curve changes due to attacks aim for disturbing system operation}. This unexpected oscillation is likely larger than the system can sustain, and lead to a serious operational issue.
	
In the scenarios described above, we assume that attackers are able to manipulate the average bid price and quantity. Maintaining this assumption, these threats can be mitigated through the use of multiple solvers and a distributed consensus algorithm. If we use multiple solvers, an attacker will need to modify the bids received by all solvers. The decentralized market option presented previously in Section~\ref{decentralized-market} provides this feature, and as shown in our prior work \cite{Eise1807:SolidWorx}, it is resistant to solver failure.  Transitioning to a decentralized market helps us to mitigate the threats presented here, but it also introduces additional challenges. TRANSAX addresses problems associated with faults in the system; however, its resistance to security threats needs further analysis. Regarding the issue of bid manipulation, a potential solution is assigning reputation values to actors, which enables removing misbehaving actors from the system. An alternative solution is imposing enforceable fines (e.g., by requiring security deposits) to disincentivize malicious or dishonest behavior. These questions can be addressed in detail as part of future work.
	
	\section{Conclusions}
	
	
In this paper, we have introduced a TES testbed, called TESST, which can simulate the operation of both centralized and decentralized TES platform. The centralized trading platform can establish all prosumers within the physical system, but it is less resilient to cyber attack. The decentralized trading platform is designed using blockchain, and it addresses security issues inherent in centralized systems. The operation and cyber-vulnerability of the centralized trading platform is analyzed in the simulation. Through simulation results, we demonstrated that it is relatively easy to manipulate a traditional centralized trading system and to cause financial and possible operational issues. The decentralized trading platform can effectively clear the market, and provide improvements to a centralized solution, but still need to be investigated using TESST in the future. This allows for the development of a secure and resilient decentralized trading platform, which is critical for the operation of TES.
	
	\Aron{We should probably add a couple more references since we have space. We can take some from, e.g., the TRANSAX paper.}
	
	\section{Acknowledgement}
	Work reported in this paper is partially supported by the ARPA-E RIAPS, Siemens CT, National Science Foundation Activity-aware Cyber-Physical Systems and the Department of Energy under Award Number DE-IA0000025 for UI-ASSIST Project. We also acknowledge help from Dr. A. Hahn, K. Kaur and the Pacific Northwest National Lab in supporting this work.
	
	\bibliographystyle{IEEEtran}
	\bibliography{references}
	
\end{document}